\documentclass[aps,preprint,preprintnumbers,amsmath,amssymb,nofootinbib]{revtex4}
\usepackage{amssymb}
\usepackage{amsmath}
\usepackage{bm}
\usepackage{graphicx}
\usepackage{epsfig}
\begin{document}
\title{Gravitational confinement of photons and matter from Induced Matter theory.}
\author{$^{1,2}$ Mauricio Bellini \footnote{ E-mail address:
mbellini@mdp.edu.ar, mbellini@conicet.gov.ar} } \vskip .2cm
\address{$^1$ Departamento de F\'isica, Facultad de Ciencias Exactas y
Naturales, Universidad Nacional de Mar del Plata, Funes 3350,
C.P. 7600, Mar del Plata, Argentina.\\  \\
$^2$ Instituto de Investigaciones F\'{\i}sicas de Mar del Plata (IFIMAR), \\
Consejo acional de Investigaciones Cient\'ificas y T\'ecnicas
(CONICET), Argentina.}

\begin{abstract}
Using an Extended Scharzschild-Anti de Sitter (ESAdS) metric here
introduced, where the noncompact extra dimension is time-like, we
study the gravitational confinement of matter and photons from a
5D vacuum. The important result here obtained is that the photon
trajectories (in general, massless particles), are periodic. They
are confined to a radius $r\leq 2 m G/c^2$ by the gravitational
field. Massive test particles are also gravitationally confined,
but their trajectories are not periodic.
\end{abstract}
\maketitle

\section{Introduction and motivation}

The possibility that gravity may have a significant role at very
short distances has attracted significant interest for many
years\cite{11,12,13} as well as the feasibility of developing a
unified gauge theory of gravitational and strong forces\cite{11}
have been discussed. In the framework of brane theories with
non-compact extra dimensions it is postulated that particles and
fields of the standard model are confined to the brane universe,
while the graviton is assumed to propagate in the bulk. Within the
classical framework of this scenario, the confining of a test
particle to the brane eliminates the effects of extra dimensions
rendering them undetectable. In general, non-gravitational forces
acting in the bulk (and orthogonal to the brane), are needed in
order to keep the test particles moving on the brane, the source
of these confining forces being interpreted in different manners.
If the notion of confinement must appear in any reasonable theory
with non-compact extra dimensions non-gravitational forces cannot
be excluded a priori. Confinement due to oscillatory behavior has
been proved in five-dimensional relativity with two
times\cite{1,2}. Indeed, null paths of massless particles in 5D
geodesic motion can appear in 4D as time-like paths of massive
particles which undergo oscillations in the 5D dimension around
the 4D hypersurface. Although the gravitational confinement has
been studied many years ago in the framework of Kaluza-Klein
theory\cite{ruth}, this issue remains unexplored in the framework
of the Induced Matter theory of gravity\cite{we}, which is
mathematically based on the Campbell-Magaard
theorem\cite{campbell,c1,campbellb,campbellc,campbelld}.

In a previous work\cite{mb} we have considered a 5D extension of
General Relativity such that the effective 4D gravitational
dynamics has a vacuum dominated equation of state: $\omega =-1$.
The starting 5D Ricci-flat metric $g_{ab}$ is determined by the
line element\cite{mb,rom}
\begin{equation}\label{a1}
dS^{2}=\left(\frac{\psi}{\psi_0} \right)^{2}\left[ c^{2}f(r)dt^{2}
- \frac{dr^2}{f(r)}-r^{2}\left(d\theta ^{2}+sin^{2}(\theta)
\,d\phi ^{2}\right)\right]- \epsilon d\psi^{2},
\end{equation}
where $f(r)=1-(2G\zeta\psi_{0}/(rc^2))[1+ \epsilon
c^{2}r^{3}/(2G\zeta\psi_{0}^{3})]$ is a dimensionless function,
$\lbrace t,r,\theta,\phi\rbrace$ are the usual local spacetime
spherical coordinates employed in general relativity and $\psi$ is
the space-like ($\epsilon =1$) extra dimension that following the
approach of the Induced Matter theory, will be considered as
non-compact. Furthermore, $\psi$ and $r$ have length units,
meanwhile $\theta$ and $\phi$ are angular coordinates, $t$ is a
time-like coordinate and $c$ denotes the speed of light. The
effective 4D metric (\ref{a1}) with $\epsilon=1$, is static,
exterior and describes spherically symmetric matter (ordinary
matter, dark matter and dark energy) on scales $r_0 < r_{Sch} <
c/H_0$ for black holes or $r_{Sch} < r < c/H_0$ for ordinary stars
with $r_0$ being the radius of the star. Furthermore, this metric
describes both, gravity (for $r < r_{ga}$) and antigravity (for $r
> r_{ga}$), $r_{ga}$ being the radius for which the effective
4D gravitational acceleration becomes zero\cite{mb}.

In this letter we shall study a new Ricci-flat metric where the
extra coordinate is time-like ($\epsilon=-1$). In this case the
metric is an Extended Schwarzschild-Anti de Sitter one (ESAdS). We
shall consider that $\psi_0$ is an arbitrary constant with length
units and the constant parameter $\zeta$ has units of
$(mass)(length)^{-1}$. The metric (\ref{a1}) is valid for
$f(r)>0$, so that its range of validity is
\begin{equation}\label{ua}
0 < r < \frac{3^{1/3} \psi_0}{3c} \frac{ \left[ \left[ c \left( 9
\xi G + \sqrt{3} \sqrt{c^4 + 27 G^2 \xi^2}\right) \right]^{2/3} -
c^2 3^{1/3}\right] }{\left[ c \left( 9 G\xi + \sqrt{3} \sqrt{c^4
+27 G^2 \xi^2} \right)\right]^{1/3} },
\end{equation}
for $\xi\neq 0$.

\section{Basic 5D equations}

For a massive test particles of a spacetime with spherical
symmetry in a 5D bulk described by (\ref{a1}), the 5D Lagrangian
can be written as\footnote{In this letter $a,b$ run from $0$ to
$4$ and Greek letters run from $0$ to $3$.}
\begin{equation}\label{b1}
^{(5)}L=\frac{1}{2}g_{ab}U^{a}U^{b}=\frac{1}{2}\left(\frac{\psi}{\psi_0}\right)^{2}\left[c^{2}f(r)
\left(U^t\right)^2-\frac{\left(U^r\right)^2}{f(r)}-r^{2}\left(U^{\theta}\right)^2-r^{2}sin^{2}
\theta\left(U^{\phi}\right)^2\right] -
\frac{\epsilon}{2}\left(U^{\psi}\right)^2.
\end{equation}
We shall take $\theta=\pi/2$. Since $t$ and $\phi$ are cyclic
coordinates, their related momentums, $p_{t}$ and $p_{\phi}$, are
constants of motion
\begin{eqnarray}\label{b2}
p_{t}\equiv\frac{\partial\,^{(5)}L}{\partial U^t}&=&c^{2}\left(\frac{\psi}{\psi_0}\right)^{2}f(r)U^t,\\
\label{b3} p_{\phi}\equiv\frac{\partial\,^{(5)}L}{\partial
U^{\phi}}&=&-\left(\frac{\psi}{\psi_0}\right)^{2}r^{2} U^{\phi}.
\end{eqnarray}
Using the constants of motion given by (\ref{b2}) and (\ref{b3}),
we can express the five-velocity condition, $g_{ab} U^a U^b =- n
c^2$, as follows:
\begin{equation}\label{b5}
\left(\frac{\psi_0}{\psi}\right)^{2}\frac{p_{t}^{2}}{c^{2}f(r)}-\left(\frac{\psi}{\psi_0}\right)^{2}\frac{\left(
U^{r}\right)^{2}}{f(r)}-\frac{p_{\phi}^2}{r^2}\left(\frac{\psi_0}{\psi}\right)^{2}-
\epsilon\left( U^{\psi}\right)^{2}= -n\,c^{2},
\end{equation}
where $n$ can take respectively the values $n=0,1$ for photons and
massive test particles. Using the expression for $f(r)$ in Eq.
(\ref{b5}), we obtain
\begin{equation}\label{b6}
\frac{1}{2}\left(
U^{r}\right)^{2}+\frac{\epsilon}{2}\left(\frac{\psi_0}{\psi}\right)^{2}\left(
U^{\psi}\right)^{2} + V_{eff}(r) = E_n.
\end{equation}
If we identify the energy $E_n$ as
\begin{equation}\label{b8}
E_n=\frac{1}{2}\left(\frac{\psi_0}{\psi}\right)^{4}(p_{t}^{2}c^{-2}+
\epsilon p_{\phi}^{2}\psi_{0}^{-2})+ \frac{n
c^2}{2}\left(\frac{\psi_0}{\psi}\right)^{2},
\end{equation}
the effective 5D potential $V_{eff}(r)$ results to be
\begin{eqnarray}
V_{eff}(r)&=&-\left(\frac{\psi_0}{\psi}\right)^{2}\frac{G\zeta\psi
_0}{r}+\left(\frac{\psi_0}{\psi}\right)^{4}\left[\frac{p_{\phi}^2}{2r^2}-\frac{G\zeta\psi
_0 p_{\phi}^2}{c^{2}r^3}\right] \nonumber \\
&-&
\frac{1}{2}\left(\frac{\psi_0}{\psi}\right)^{2}\left[\left(U^{\psi}\right)^2\left(\frac{2G\zeta\psi_0}{c^2
r}+\epsilon\frac{r^2}{\psi_0^2}\right)+\epsilon
\left(\frac{rc}{\psi_0}\right)^{2}\right]. \label{b7}
\end{eqnarray}
In order to study the effective 4D manifestation of the potential
(\ref{b7}) for both, massive particles and photons, we shall study
a static foliation $\psi=\psi_0=c/H_0$, such that the dynamics
evolves on an effective 4D manifold $\Sigma_0$. From the point of
view of an relativistic observer, this implies that $U^{\psi}=0$.

\section{Effective 4D dynamics}

In this section we shall study the effective 4D dynamics on the
hypersurface $\psi=\psi_0=c/H_0$, which can be obtained by
considering a frame $U^{\psi}=0$. This fact implies from the point
of view of the geodesic trajectories that they are described by
the equation
\begin{equation}\label{geo}
\frac{dU^a}{dS} +\Gamma^a_{\,\,bc}\,U^b U^c=\phi^a,
\end{equation}
where $U^a={dX^a\over dS}$ and $\phi^a$ is an external force.
Furthermore, the velocity condition (\ref{b5}) must be fulfilled.

From the geodesic point of view, the equation (\ref{geo}) implies
that
\begin{eqnarray}
&& \frac{dU^{\alpha}}{dS} +\Gamma^{\alpha}_{\,\,a b}\,U^{a}
U^{b}=\phi^{\alpha},\\
&& \frac{dU^{\psi}}{dS} +\Gamma^{\psi}_{\,\,a b}\,U^{a}
U^{b}=\phi^{\psi}_{(n)},
\end{eqnarray}
where
\begin{eqnarray}
&&  \phi^{\alpha} = 0, \label{10} \\
&& \phi^{\psi}_{(n)}=\frac{n}{\psi_0}. \label{force} \label{conf}
\end{eqnarray}
In the eq. (\ref{10}) we have supposed the non existence of an
additional fifth force: $\phi^{\alpha}=0$. The only non-zero force
that we shall consider is $\phi^{\psi}_{(n)}$, but only for
massive test particles. In this case it plays the role of a
confining force.

\subsection{Photon trajectories on the brane}

\begin{figure*}
\includegraphics[height=15cm]{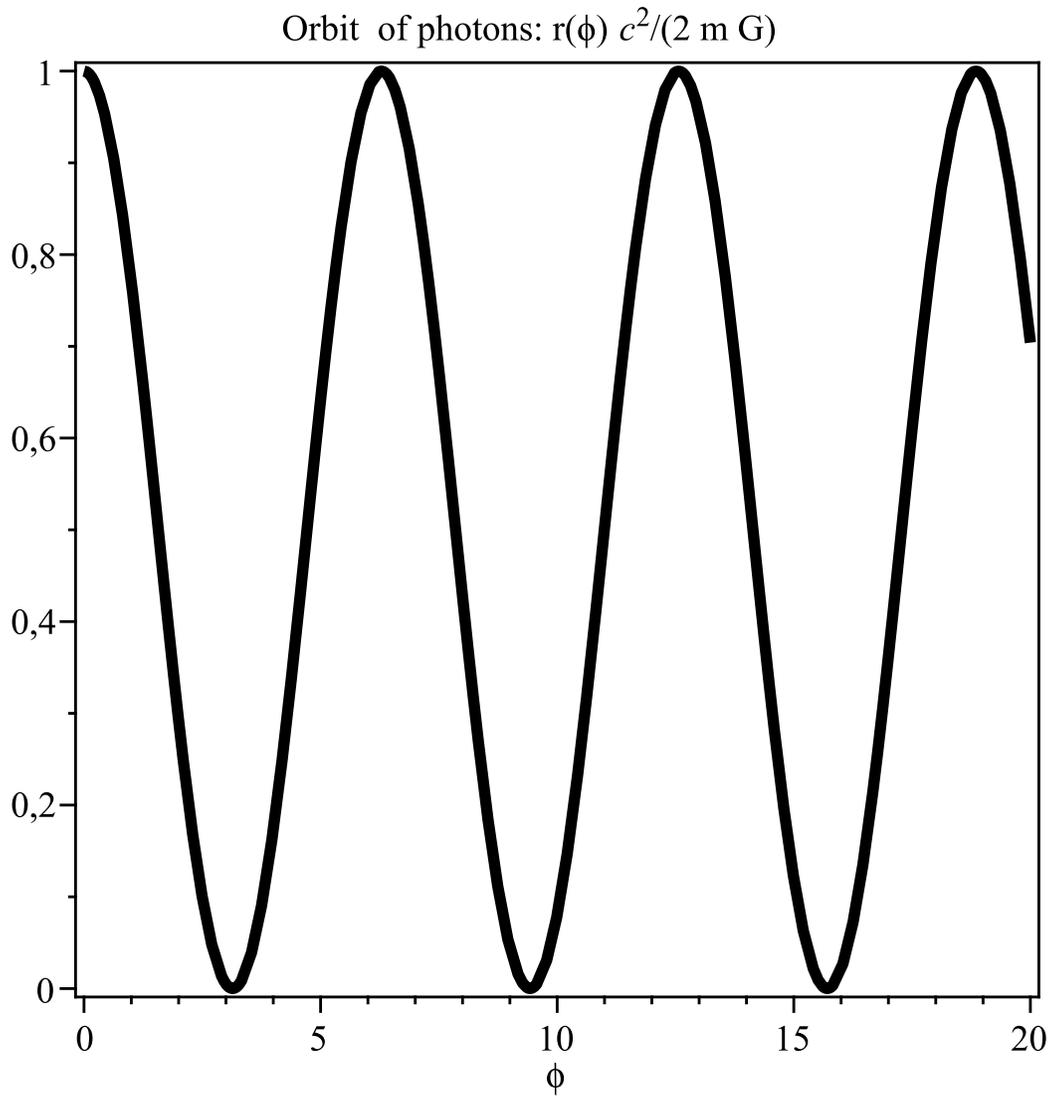}\caption{\label{figura1}
Orbit of photons $r(\phi)c^2/\left({2 m G}\right)$. Notice that
the periodic trajectories remain confined to $ r \leq 2 m G/c^2$.}
\end{figure*}

In order to consider the orbital equation for the photons, we take
$u(\phi) = 1/r(\phi)$ and using (\ref{b2}) and (\ref{b3}) in Eq.
(\ref{b5}), we obtain the following equation for the orbits of
photons on the brane:
\begin{equation}
\frac{d^2u}{d\phi^2} + u \left[ 1 - \left(\frac{3 m G}{c^2}\right)
\, u \right] = 0,
\end{equation}
where we have obtained the 4D hypersurface after taking the
foliation $\psi=\psi_0$. A particular solution for this equation
is $u(\phi) = c^2 {\left[ \tan^2{\left(\phi/2\right)}
+1\right]\over 2 m G}$, so that
\begin{equation}\label{solutions}
r(\phi) = \frac{2 m G}{c^2 \, \left[ \tan^2{\left(\phi/2\right)}
+1\right]}.
\end{equation}
In the figure \ref{figura1} we have plotted the radius as a
function of $\phi$: $r(\phi)$. For simplicity we have taken ${2 m
G\over c^2}$. Notice that the trajectories are closed so that
photons remain on radius on the range $0 < r(\phi) < {2 m G\over
c^2}$.

\subsection{Test massive particles}

If we consider test massive particles, we obtain from eqs.
(\ref{b2}) and (\ref{b3}) that the momentums $\bar{p}_t$ and
$\bar{p}_{\phi}$ on the four dimensional hypersurface $\Sigma_0$,
are
\begin{eqnarray}\label{bb2}
\bar{p}_{t}\equiv\left.\frac{\partial\,^{(5)}L}{\partial U^t}\right|_{\psi=\psi_0}&=&
\left. c^{2}\left(\frac{\psi}{\psi_0}\right)^{2}f(r)U^t\right|_{\psi=\psi_0},\\
\label{bb3} \bar{p}_{\phi}\equiv \left.
\frac{\partial\,^{(5)}L}{\partial U^{\phi}}\right|_{\psi=\psi_0}
&=&- \left.\left(\frac{\psi}{\psi_0}\right)^{2}r^{2}
U^{\phi}\right|_{\psi=\psi_0}.
\end{eqnarray}
Because we are dealing with a 5D static metric on which we make a
static foliation $\psi=\psi_0$, we choose $U^{\psi}=0$. Therefore,
if we take this foliation in the potential (\ref{b7}), we obtain
the effective 4D potential on the brane
\begin{equation} \bar{V}_{eff}(r) = - \frac{m G}{r} +
\bar{p}^2_{\phi} \left[ \frac{1}{2 r^2} - \frac{m G}{c^2 r^3}
\right] + \frac{c^2}{2} \frac{r^2}{\psi^2_0}.
\end{equation}
In the absence of angular moments, i.e. for $\bar{p}_{\phi}=0$,
one obtains only gravitational effects, so that the effective 4D
gravitational force is
\begin{equation}
\vec{a} =  - \vec{\nabla} \bar{V}_{eff}(r) = - \left[\frac{m
G}{r^2} + r H^2_0 \right] \hat{r} <0, \label{force}
\end{equation}
which is related to the effective 4D gravitational potential with
$\bar{p}_{\phi} =0$ and $m=\xi c/H_0$: $\bar{V}_{eff}(r)$
\begin{equation}
\bar{V}_{eff}(r) = \frac{1}{2} r^2 H^2_0 - \frac{G m}{r}.
\label{bb}
\end{equation}
In other words, $\xi$ is a parameter that can be obtained from the
size of the horizon $\psi_0=c/H_0$ and the mass of the source $m$.
Notice that the central acceleration is always attractive for any
massive source. However its value takes a point of inflection at
$r=r_*$, where $\left.\bar{V}_{eff}(r)\right|_{r_*} =0$, being
\begin{equation}
r_* = \left(\frac{2 m G}{H^2_0}\right)^{1/3}.
\end{equation}
For values $r> r_*$ the gravitational potential (\ref{bb}) is
positive and there is no confinement.

The effective 4D Einstein equations being given by by
\begin{equation}
\bar{G}^{\alpha}_{\beta} = -8\pi G \,\bar{T}^{\alpha}_{\beta},
\end{equation}
such that the effective 4D energy-momentum tensor in a static
frame [$u^{\alpha} = e^{\alpha}_{\,\,a} U^{a}$]
\begin{equation}
\bar{u}^t = \bar{u}^{\phi} = \bar{u}^{\theta}=0, \qquad \bar{u}^r
= \pm \sqrt{f(r)} c,
\end{equation}
is
\begin{equation}
\bar{T}^{\alpha}_{\,\,\beta} = \left( P+\rho\right)
\bar{u}^{\alpha} \bar{u}_{\beta} - \delta^{\alpha}_{\,\,\beta}
\,\rho.
\end{equation}
Since the effective 4D relevant components of the Einstein tensor
are
\begin{equation}
\bar{G}^0_{\,\,0} = \bar{G}^r_{\,\,r}
=\left.\frac{3}{\psi^2_0}\right|_{\psi_0=c/H_0} = \frac{3
H^2_0}{c^2},
\end{equation}
we obtain that the effective 4D equation of state for the induced
gravitational system is $\omega = P/\rho = -1$.

\subsection{Trajectories of massive test particles}

When we go down from 5D to 4D the differential equation for the
orbits of massive test particles is
\begin{equation}\label{b13}
\frac{d^{2}r}{d\phi^2}+ \frac{r^4}{\bar{p}_{\phi}^{2}}
\,\frac{d}{dr}\left[\bar{V}_{eff}(r)\right]+\frac{4
r^3}{\bar{p}_{\phi}^2}\bar{V}_{eff}(r)
-2\left[\frac{c^2}{\bar{p}_{\phi}^2}+\frac{
\bar{p}_{t}^2}{c^2\bar{p}_{\phi}^2}-\frac{c^2}{H_0^2}\right]
r^3=0,
\end{equation}
where $\bar{V}_{eff}(r)$ is given by (\ref{bb}). The general
solution for this equation is
\begin{eqnarray}
\phi - \phi_0 &= & \pm \int^{s=r(\phi)} \frac{6 \psi_0
\bar{p}_{\phi} c \,\, ds}{\sqrt{36 C_1 \bar{p}_{\phi}^2 c^2
\psi^2_0 - 120 m G \psi^2_0 s^3 - 60 c^4 s^6 + 18 {\rm C}
\bar{p}_{\phi}^2 c^2 \psi^2_0 s^4}},
\end{eqnarray}
where ${\rm C}= c^2 \left({1\over \bar{p}_{\phi}^2} - {1\over
H^2_0} \right) + {1\over c^2} {\bar{p}_{t}^2\over
\bar{p}_{\phi}^2}>0$. One must require that $36 C_1
\bar{p}_{\phi}^2 c^2 \psi^2_0 - 120 m G \psi^2_0 r^3 - 60 c^4 r^6
+ 18 C \bar{p}_{\phi}^2 c^2 \psi^2_0 r^4
>0$, in order to the solution be real. For the special case where
$C_1=0$, this condition implies that the radius of any trajectory
for one massive test particle cannot exceed the value
\begin{eqnarray}\label{ie}
0 &< & r < \frac{\left[ 10 \sqrt{10000 G^2 m^2 c^4 - 10
\frac{c^4}{H^2_0} \left(1- \frac{\bar{p}^2_{\phi}}{H^2_0} \right)
+ \frac{\bar{p}^2_t}{H^2_0}}-1000 G m c^2\right]^{1/3}}{10 c^2}
\nonumber \\
& + & \frac{\left[ \frac{c^4}{H^4_0} \left( 1 -
\frac{\bar{p}^2_{\phi}}{H^2_0}\right) + \frac{\bar{p}^2_t}{H^2_0}
\right]}{\left[ 10 c^2\sqrt{10000 G^2 m^2 c^4 - 10
\frac{c^4}{H^2_0} \left(1- \frac{\bar{p}^2_{\phi}}{H^2_0} \right)
+ \frac{\bar{p}^2_t}{H^2_0}}-1000 Gmc^2\right]^{1/3}},
\end{eqnarray}
where, in order to ${\rm C} >0$, we must require that
\begin{equation}\label{iee}
\left( 1 +\frac{\bar{p}^2_t}{c^2} \right) >
\frac{\bar{p}^2_{\phi}}{H^2_0}.
\end{equation}
All possible trajectories for massive test particles are given by
the solutions of eq. (\ref{b13}) [with the conditions (\ref{ie})
and (\ref{iee}) included], and can be obtained by numerical
methods. As can be noted, they are not periodic. However this
issue goes beyond the scope of this letter.

\section{Final Comments}

We have introduced a new Ricci-flat 5D static ESAdS metric such
that the extra coordinate is time-like and noncompact. This metric
is very interesting because, after make a static foliation
$\psi=\psi_0=c/H_0$ on the extra coordinate, we obtain an
effective 4D hypersurface which describes the gravitational
confinement of the massive test particles on a radius given by the
condition (\ref{ua}). Furthermore, the effective 4D equation of
state for the induced gravitational system is also $\omega =
P/\rho = -1$, as in the case with $\epsilon=1$. Notice that in the
eq. (\ref{10}) the only non-zero force that we shall consider for
massive test particles is $\phi^{\psi}_{(n)}$, which plays the
role of a confining force on the 4D hypersurface obtained after
making the static foliation. However, for massless particles like
photons this force is null. In this case massless particles
exhibit periodic orbits, with radius $ r(\phi) \leq (2 m G)/c^2$.
This can be seen in the Fig. (\ref{figura1}), where we have
plotted the particular solutions (\ref{solutions}), which describe
the periodic orbit of photons for arbitrary initial conditions. Of
course, this kind of solutions for photons demonstrates that they
are confined by the gravitational field with a metric (\ref{a1}),
in which $\epsilon=-1$.

\section*{Acknowledgements}

\noindent M.B. acknowledges UNMdP and CONICET Argentina for
financial support.

\end{document}